# Focusing through scattering medium: a fundamental trade-off between speckle size and intensity enhancement

Eitan Edrei and Giuliano Scarcelli*

*Fischell Department of Bioengineering, University of Maryland,*
*8278 Paint Branch Drive, College Park, MD 20742, USA*
*E-mail address: scarc@umd.edu*

Focusing light through highly scattering materials by modifying the phase profile of the illuminating beam has attracted a great deal of attention in the past decade paving the way towards novel applications. Here we discovered a tradeoff between two seemingly independent quantities of critical importance in the focusing process: the size of the focal point obtained behind a scattering medium and the maximum achievable brightness of such focal point. We theoretically derive and experimentally demonstrate the fundamental limits of intensity enhancement of the focal point and relate them to the intrinsic properties of the scattering phenomenon. We demonstrate that the intensity enhancement limitation becomes dominant when the focusing plane gets closer to the scattering layer thus limiting the ability to obtain tight focusing at high contrast, which has direct relevance for the many applications exploring scattering materials as a platform for high resolution focusing and imaging.

Light-based imaging and focusing methods have been historically limited to transparent materials or shallow depths due to multiple light scattering in complex media [1]. Traditional methods to combat aberrations and distortions by measuring and projecting complementary phase maps (e.g. adaptive optics) [2, 3] have been able to compensate for mild aberrations due to imperfect optical elements, atmospheric turbulence and distortions within the eye [3-5]. Yet, they are largely ineffective in highly scattering media due to the numerous amount of degrees of freedom involved and the short scattering mean-free-path. Starting with the pioneering work by Vellekoop and Mosk [6], the past decade has seen tremendous progress in our ability to focus a laser beam through a highly scattering material. In this process the "focal point" is obtained by aligning the relative phases of light emerging from the scattering medium to constructively interfere at a point of interest. This can be achieved either by iteratively modifying the incident beam phase profile with a spatial light modulator (SLM) [6-8], by directly measuring the optical transmission matrix of the scattering medium [9-11] or by recording the field fluctuations induced by the medium [12, 13]. The intriguing ability to deliver light through disordered materials has attracted a great deal of interest for diverse applications such as deep-tissue focusing [14], optogenetic modulations [15], imaging of hidden objects [16, 17] and high resolution focusing/microscopy [18-20].

The underlying concept shared by these innovative works is that the combination of a scattering medium with spatially-resolved control of the light beam phase profile can effectively work as a lens. Several enabling features of such "scattering lens" systems have been described such as super-resolution focusing [21], versatile focal length and structural compactness [22, 23]. However, the fundamental features and limitations of the focusing capabilities of "scattering lenses" are not fully understood. Addressing this question, here we discovered a fundamental tradeoff between the size of the smallest speckle (serving as focal point) that can be obtained behind a scattering medium and the brightness of such focal point achieved via intensity enhancement. We present a theoretical derivation and experimental demonstration that as the focal plane gets closer to the scattering material leading to smaller speckle size, the intensity enhancement of the focal point within this plane is severely compromised. We show that this fundamental limit imposes practical constraints on focusing protocols, as it effectively limits the size of a focal point enhanced through a scattering layer, and/or sets an upper-bound to the intensity flux delivered to a given location within a scattering medium.

The intensity enhancement at a point behind a scattering layer (i.e. the ratio between optimized focus intensity and average background) for a monochromatic coherent light has been previously described as [8]:

$$I_{enhancement} = \gamma N \qquad (1)$$

where $N$ is the number of controllable degrees of freedom to modify the phase profile of the illuminating beam and $\gamma$ is an experimental scaling factor. For polychromatic light sources the enhancement will be reduced proportionally to the number of transmitted independent frequency components [24-27]. The enhancement in Eq. (1) can be understood intuitively as the result of adjusting the phases (e.g. via an SLM) of $N$ independent sub-sources within the beam so that they constructively interfere at a desired location; the pre-factor $\gamma$ depends on several experimental parameters such as the operation mode of the SLM, the sensitivity of the camera to small intensity changes, the noise level throughout the enhancement process and the stability of the scattering medium [7, 28, 29].

Here we find that the intensity enhancement is not generally constant when focusing light behind a scattering medium and that Eq. (1) represents the upper limit of intensity enhancement that can be reached. We derive a general expression for intensity enhancement in terms of the fundamental characteristics of the scattering phenomenon (e.g. scattering divergence angle, density of scattering elements, beam spatial coherence). Importantly, we find that these scattering properties introduce severe limitations as the focusing plane gets closer to the scattering layer thus compromising the ability to enhance the intensity of a focal point when high resolution is desired.

To quantitatively derive the focusing limits, we consider the general scenario to achieve an enhanced focal point $P$ at a plane located a distance $z$ from a scattering layer. To optimize the constructive interference at point $P$; a phase map of linear dimension $D$ is projected by an SLM onto the scattering layer and is optimized using a continuous sequential algorithm [30] (Fig. 1).

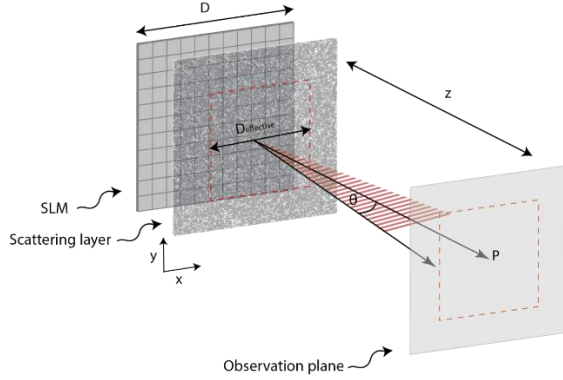

FIG. 1. Schematics of a general procedure to form a focal point at location $P$ in a plane displaced by the axial distant $z$. The SLM surface is imaged on a scattering layer to spatially control the phase profile of the incident beam.

Because the scattering process is characterized by a divergence angle $\theta$, light diffused by the scattering layer will not be re-directed everywhere; instead, each scattering event will re-direct light within a cone of angle $\theta$ (Fig. 1). Locations on the scattering layer for which the scattering cone does not include point $P$ will not contribute to the enhancement process. As a result, only a portion of the illuminated area will contribute to the intensity enhancement. Accordingly, for every experimental scenario, we can define an effective area $A_{effective}$ which contributes to the intensity at point $P$. To quantify $A_{effective}$, we model the illumination plane as a collection of scattering point sources and assign a weight factor to each point based on their effective contribution to the intensity of focal point $P$. Specifically, we can assume that each scattering point source generates a beam with a transverse gaussian profile of width determined by the scattering angle $\theta$ and by the propagation distance $z$, so that the standard deviation of the gaussian is $z \cdot tan(\theta)$. The effective area is the weighted integral of all the scattering point sources as follows:

$$A_{effective} = \int_0^D \int_0^D e^{\frac{-(x^2+y^2)}{2(z \cdot tan\theta)^2}} dxdy \quad (2)$$

where $x, y$ are the spatial coordinates in the plane of the scattering layer (i.e: $z = 0$).

In the limit of large distances from the scattering layer, i.e. for $z \to \infty$, the integration of Eq. (2) yields $A_{effective} = D^2$, i.e. all SLM pixels equally contribute to the optimization process. This is the ideal situation described by Eq. 1.

For finite distances from the scattering layer, the integration of Eq. (2) can be solved analytically by substitution, $\tilde{x}(\tilde{y}) = x(y) \cdot \frac{1}{\sqrt{2} \cdot z tan(\theta)}$, to yield an effective illumination area of:

$$A_{effective} = \frac{\pi}{2} \cdot (z \cdot tan\theta)^2 \cdot \left[erf(\frac{D}{\sqrt{2} \cdot z \cdot tan(\theta)})\right]^2 \quad (3)$$

Thus, only SLM pixels that fall within the effective illumination area will contribute to the enhancement. This leads to the general form of Eq. (1) which includes the underlying physics of the scattering phenomenon:

$$I_{enhancement} = \gamma N \cdot \frac{A_{effective}}{A}$$
$$= \gamma N \cdot \frac{\pi}{2D^2} \cdot (z \cdot tan\theta)^2 \cdot \left[erf(\frac{D}{\sqrt{2} \cdot z \cdot tan(\theta)})\right]^2 \quad (4)$$

In summary, the ideal enhancement would be reached in perfectly isotropic scattering conditions where the light is distributed equally over a solid angle of $2\pi$ after the scattering medium. Instead, even though the intensity distribution of the SLM pattern projected onto the scattering layer is uniform, the finite divergence angle typical of a scattering layer introduces a weighing function that assigns smaller contributions to peripheral locations. The intensity enhancement can thus be interpreted as arising from a radially degrading intensity distribution, a scenario which has a direct impact on both the focal intensity and the effective numerical aperture as we discuss next.

It is interesting to analyze the behavior of the intensity enhancement as a function of the unitless parameter: $U = \frac{D}{\sqrt{2} \cdot z \cdot tan(\theta)}$:

$$I_{enhancement} = \gamma N \cdot \frac{\pi}{4} \cdot \left(\frac{1}{U}\right)^2 \cdot [erf(U)]^2 \quad (5)$$

We note two limits: $\lim_{U \to \infty} \left[\frac{\pi}{4} \cdot \left(\frac{1}{U}\right)^2 \cdot [erf(U)]^2\right] = 0$, $\lim_{U \to 0} \left[\frac{\pi}{4} \cdot \left(\frac{1}{U}\right)^2 \cdot [erf(U)]^2\right] = 1$. The first limit occurs for focal planes very close to the scattering layer, i.e. $z \ll D$. Under these circumstances, the effective illumination area vanishes and the enhancement approaches zero (note though that our derivation does not consider the evanescent field and thus is restricted to the regime $z > \lambda$). The second limit refers to the situation where the focal plane is far from the scattering layer. In this case, the effective illumination area is the entire illumination area, and the enhancement approaches the optimal $\gamma N$ value. Interestingly, this limit can be expressed as: $\frac{D}{\sqrt{2} \cdot z \cdot tan(\theta)} \ll 1$ and reduces to

$$z \gg \frac{d \cdot D}{\lambda} \quad (6)$$

where $d$ is the average linear distance between scattering particles and we approximated the scattering angle as $tan(\theta) \approx \theta \approx \frac{\lambda}{d}$ [31] (valid under the condition $d > \lambda$). This limit exactly coincides with the 'far-field' condition for partial coherent light: $z \gg \frac{\Delta\mu \cdot D}{\lambda}$ derived from the propagation of mutual coherence as described by the generalized Van Cittert- Zernike theorem [32-34], where $\Delta\mu$ is the coherence length right after the scattering material, which was shown to approach $d$ [31]. This relation thus links the maximum intensity enhancement achievable to fundamental properties of the scattering phenomenon such as the concentration of scattering elements and the spatial coherence of the light beam. Interestingly, while the focal length of scattering lenses has been so far assumed to be entirely variable [19, 35], here we find that a 'far-field' condition needs to be met for optimal focusing. Our treatment is general and depends only on the illumination size and the length scale $d$, while the specific way the SLM phase map is imaged onto the scattering medium can be chosen arbitrarily.

To experimentally verify our theoretical predictions, we used the setup illustrated in Fig. 2(a). A polarized expanded laser beam of $\lambda = 660\ nm$ (LaserQuantum) was reflected off the surface of phase-only SLM (Hamamatsu X10468-01). The SLM plane was then imaged on a 600-grit diffuser (Thorlabs) which served as our scattering layer. To obtain different sizes of illumination we used a de-magnifying 4-f imaging system, with L1 of focal length 400 mm and L2 of variable focal length (45mm to 3mm). An infinitely corrected imaging system was used after the scattering layer to record a plane of interest (L3 = 0.75 NA, 20X, L4 = 200 mm). The distance between the scattering medium and the observation plane (i.e. the plane of enhancement) was selected by adjusting the translational stage of L3. To enhance a single point beyond the scattering layer, the SLM was divided into 100 macro-pixels, and each pixel was varied individually from 0 to $3\pi$ to determine the optimal phase configuration by using the recorded pattern on the camera as a feedback [7]. This process was repeated for all pixels twice, yielding a total time $\sim 20min$ for a single enhancement process.

FIG. 2. (a) Experimental setup: the SLM surface (divided into 100 macro-pixels) was imaged via a 4-f imaging system (L1, L2) onto the scattering layer. A plane behind the scattering layer was imaged onto the camera, and a location within the imaged plane was enhanced by applying a feedback loop to vary the phase map of the SLM. The imaged plane was selected by adjusting the translational stage of L3. (b) Intensity distribution pattern before the enhancement process. (c) Intensity distribution after the enhancement process (scale bar = 5µm). The inset shows the final SLM phase map to which the algorithm converged.

First, we verified our ability to focus light through a scattering material consistently with traditional protocols. We imaged a plane located 1.6 mm after the scattering layer (corresponding to $z \sim 3.2D$) and as expected obtained a speckle pattern, shown in Fig. 2(b). We selected a central location from the recorded pattern and sequentially varied each pixel of the SLM to enhance the intensity recorded at that location. After two iterations of every pixel, we arrived at the final intensity distribution presented in Fig. 2(c). We reached an enhancement of 32 corresponding to $\gamma \sim 0.3$, consistent with previously reported values [8, 29]. Note, that the number of degrees of freedom controlled by the SLM is orders of magnitude smaller than that needed to perfectly correct for the variations of the scattering medium. Yet, by adjusting the relative phases to constructively interfere at the desired location, a small portion of the light energy is redirected to form a high contrast intense focal point.

Next, we directly demonstrated the prediction of Eq. (4). We de-magnified the SLM onto the scattering layer to an area of linear dimension $D = 500\ \mu m$, and executed the enhancement protocol at planes of different distances from the scattering layer. Figure 3(a) shows the intensity enhancement as the selected focal plane gets closer to the scattering layer (orange dots). The black line is a fit to the experimental data using Eq. (4) and keeping $\tan(\theta)$ as a free parameter. As evident from Fig. 3(a), the intensity enhancement is not constant as Eq. (1) would predict but increases with the distance from the scattering layer, in agreement with the theoretical prediction of Eq. (4).

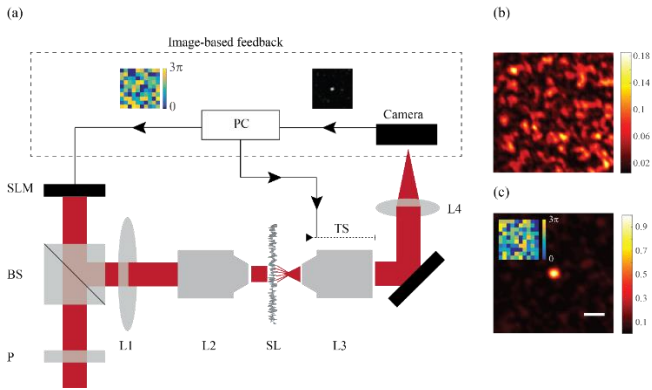

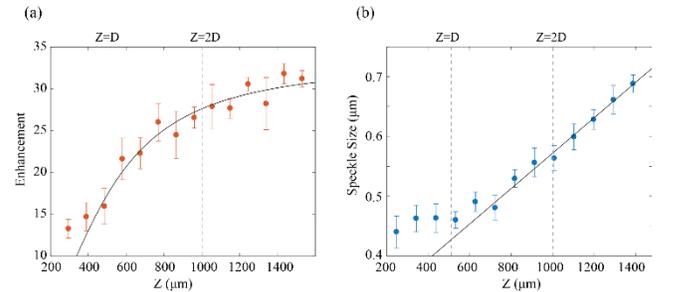

FIG. 3. (a) Intensity enhancement at different focusing planes of varying distance from the scattering layer. Experimental data (orange dots) are well fit by Eq. (4) (black line). The optimization algorithm was performed >10 times for each data point and results below the median were discarded to eliminate artifacts due to mechanical vibrations or material decorrelation. (b) Speckle size at different focusing planes of varying distance from the scattering layer. Experimental data are calculated at HWHM of the peak (blue dots). A linear fit (black line) fits well the data after the critical distance.

Figure 3(b) shows the corresponding average speckle size obtained at various planes after the scattering layer, which displays a known behavior, i.e. the speckle size is constant until a critical distance ($z_c \approx \frac{d \cdot D}{\lambda}$ [33, 36]) and then scales up linearly. Interestingly, this effect can also be explained using the effective area concept: before the "far-field" condition of Eq. 6, the effective area contributing to the constructive interference proportionally decreases and thus prevents further reduction of the speckle size [36]. Figures 3(a) and 3(b) are consistent with each other: using the value for $\theta$ obtained from the fit of Fig. 3(a), the critical axial location for linear speckle growth is $z_c \approx 1.65D$. Since in this experiment $D = 500\mu m$ this value yields $z_c \approx 800\mu m$ which corresponds well to the transition to linear speckle growth observed in Fig. 3(b).

To prove the universality of our findings, we repeated the experiments in Fig. 3 for three different sizes of illumination ($D$=250µm, 500µm, 1250µm) by varying the de-magnification of the SLM image onto the scattering layer. To compare the results, we considered that Eq. (4) reduces to $I_{enhancement} \approx 0.73\gamma N$ at the critical distance $z = z_c$ (under the approximation $tan(\theta) \approx \theta \approx \frac{\lambda}{d}$ [23]). Therefore, for each illumination size, we plot the value of $z$ corresponding to 73% enhancement of the maximal value, which should correspond to the critical distance $z_c$. The results are presented in Fig. 4(a):

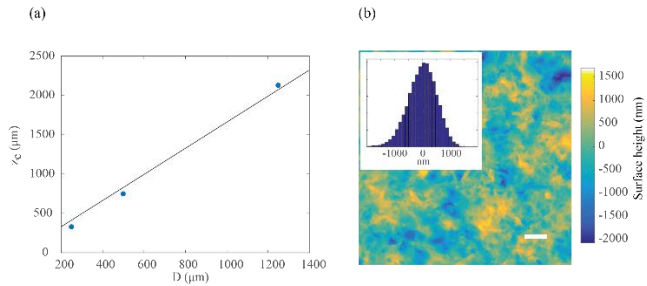

FIG. 4. (a) Critical distance $z_c$, evaluated as the axial location corresponding to 73% of the maximal intensity enhancement, vs the illumination linear dimension $D$ (blue dots). The black line is a linear fit to the data through the origin of the coordinates. (b) AFM surface height map of the scattering layer (scale bar = $10\mu m$). The inset shows the distribution of surface heights.

The correlation between $D$ and $z_c$ is expected to be governed only by the divergence angle: $\theta \sim \frac{\lambda}{d}$ and is therefore constant for any illumination size. Indeed, the data from the three illumination sizes are well described by a linear fit. From the slope of the fit, we extracted the scattering divergence angle $\theta \sim \frac{\lambda}{d} = 0.6$, corresponding to $d = \frac{\lambda}{0.6} = 1100\ nm$. We confirmed the estimate of the scattering scale $d$ by mapping the height of the diffuser with an Atomic Force Microscope (AFM) as shown in Fig. 4(b). From the AFM measurement the FWHM of the variation distribution is $\sim 1200\ nm$, in good agreement with our calculated value. In biological tissues the scattering angle is typically smaller, and the scattering is anisotropic with forward scattering being the preferred direction [37, 38].

Our analysis leads to important practical consequences regarding focusing applications through scattering materials. Let's examine the experimental scenario where we want to distinguish two-point objects (at distance $r$) and we illuminate only one of them with the enhanced focal point. In this situation, the relevant specification to optimize for imaging or focusing applications is the contrast ratio between the two points, which depends both on the intensity enhancement and on the width of the focal point. Assuming the focal point to have a Gaussian intensity profile of width $\sigma = \frac{z\lambda}{4D}$ given by the linear regime of Fig. 3(b), the contrast ratio can be evaluated combining Eq. (4) and the intrinsic contrast of the Gaussian intensity distribution:

$$C_r(z) = \gamma N \frac{\pi}{2D^2}(z \cdot tan\theta)^2 \left[ erf\left(\frac{D}{\sqrt{2} \cdot z \cdot tan(\theta)}\right) \right]^2 \left(1 - e^{-\frac{r^2}{2\sigma^2}}\right) \quad (7)$$

Fig. 5(a) illustrates the contrast ratio for two objects at distance $r$ according to Eq. (7). The distance $r$ between the two objects determines the needed resolution and, as a consequence, the focusing plane $z$ with maximal contrast ratio. For example, if a resolution of $\lambda$ is desired, the maximal contrast ratio will be achieved at a plane $z \approx 2D$. Trying to enhance a focal point in a plane closer to the scattering layer will degrade the contrast due to lack of enhancement efficiency; trying to enhance the focal point in a plane farther from the scattering layer will degrade the contrast because the focal point widens. The set of maxima of Fig. 5(a) forms the Modulation Transfer Function (MTF) of the system, shown in Fig. 5(b) as a function of $\lambda/r$, a unitless measure of the spatial frequency of interest. As in traditional imaging systems, the modulation transfer function decays at high spatial frequencies; however, unlike traditional imaging systems, the best MTF values of this imaging system also depend on the observation plane $z$.

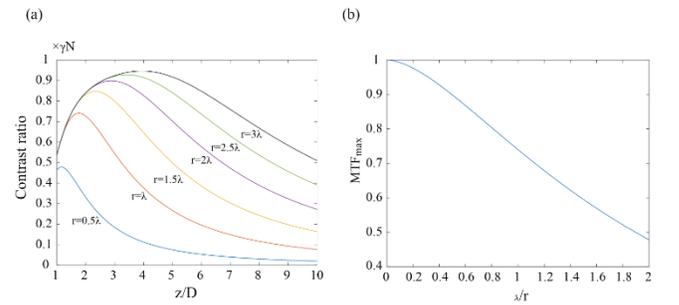

Fig. 5. (a) Contrast ratio of the enhanced focal point as a function of distance from the scattering layer for different lateral resolutions. The optimal contrast depends on the required resolution and is obtained at a specific distance from the scattering layer. (b) The MTF of the system evaluated by collecting the maxima of (a).

In summary, in this work we derived and verified the fundamental limits of intensity enhancement that can be reached when focusing light through a scattering material at different axial locations. This has direct relevance for the many studies that use scattering materials as a platform for high resolution microscopy/focusing through the generation of a sharp focal point behind the scattering layer [18-20, 39]. Our analysis provides a comprehensive framework to determine the maximum contrast achievable when high-resolution or super-resolution is attempted with scattering lenses. Optimal enhancement is achieved by imaging as many SLM pixels as possible into the effective illumination area calculated here; thus, in practice, the ultimate enhancement limit is reached when the SLM pixels are de-magnified to the smallest size allowed by the imaging system that projects the SLM map onto the scattering material. In the evanescent-wave regime ($z \sim 100\ nm$) where the size of the speckles is less than $\lambda/2$, sub-wavelength resolutions can be reached [40]. However, also in this regime, the effect of the divergence angle needs to be considered to quantify the maximum intensity enhancement. Indeed, the effective illumination area in the evanescent regime is expected to be reduced to several microns, which will limit the intensity enhancement. This could explain why experimentally the enhancement was found to be far from optimal in this regime [21]. For practical applications, it will be important to establish how much intensity enhancement is required to achieve sufficient contrast for specific purposes, such as fluorescence excitation, neural activity modulations or label-free imaging. This will ultimately determine how close the focusing plane can be set and how high resolution can be achieved. The phenomenon we describe here may also be applicable to the emerging field of focusing and imaging through multimode fibers [41-44] as the enhancement capabilities at close proximity to the fiber outlet is expected to decrease.

In conclusion, we theoretically derived and experimentally verified the fundamental limits of intensity enhancement when focusing light through a scattering material. We found that the enhancement is severely limited as the distance between the scattering layer and the focal plane is decreased while it approaches the optimal value of $\gamma N$ in the far field where the focal point, hence the resolution of the optical system, is not optimal. In this work we obtained the coherent focal point using an iterative process; however, our derivation is general and hence valid for other methods for which coherent focusing is obtained such as optical phase conjugation [12], transmission matrix inversion [9] or time reversal [45].